\documentclass{article}
       \usepackage{amsmath,amsthm,upref}
       \usepackage{amssymb,amsbsy}
       \usepackage{amsfonts}
\usepackage{cite}
\usepackage{times}

\normalfont\upshape

\begin{document}

\author{{\bf S. A. Duplij\footnote{On leave of absence from 
V.N. Karazin
Kharkov National University, Svoboda Sq. 4, Kharkov 61022, Ukraine.}}\\
{\it Center for Mathematics, Science and Education}\\
{\it Rutgers University,}
{\it 118 Frelinghuysen Rd., Piscataway, NJ 08854-8019}\\
{ duplij@math.rutgers.edu,
http://homepages.spa.umn.edu/\~{}duplij}}
 
 \title{{\bf A NEW HAMILTONIAN FORMALISM FOR SINGULAR LAGRANGIAN THEORIES} }

\date{
27 November 2011 \\
}
\maketitle
\noindent
{\small

\noindent We introduce a version of the Hamiltonian formalism based on  the
Clairaut equation theory which allows 
us a self-consistent description of systems with degenerate (or singular)
Lagrangian. 
A generalization of the Legendre transform to the case when the Hessian 
is zero is done using the mixed (envelope/general) solutions of 
the multidimensional Clairaut equation. The corresponding system of equations 
of motion
is equivalent to the initial Lagrange equations, but contains ``nondynamical'' 
momenta and unresolved velocities. This system is reduced to
the physical phase space and presented in the Hamiltonian form by introducing 
a new (non-Lie) bracket.

 \par \noindent {\bf KEY WORDS}:\hskip 3pt
Legendre transform, Hessian, 
multidimensional Clairaut equation, non-Lie algebra, Poisson brackets}
%
\bigskip
\medskip

\newpage
\section*{Introduction}
Modern gauge theories on a classical level are singular theories described by degenerate Lagrangians. Their quantization is based on
 generalized versions of the Hamiltonian formalism. The standard approach
is the Dirac constraint method  \cite{dirac}. In this way, it would  be worthwhile to investigate
other methods to present a singular theory in a Hamiltonian-like form. Our previous result \cite{dup_belg2009} was in generalizing the Legendre
transformation to singular (degenerate) Lagrangians (with zero Hessian matrix). For that a mixed (general/envelope) solution of
the multidimensional Clairaut equation was introduced \cite{dup_belg2009}.

In this paper we apply the above idea to construct a self-consistent
version of the canonical (Hamiltonian)
formalism and present an algorithm to describe any singular Lagrangian
system without introducing constraints. To simplify matters we use 
coordinates, but all the statements can be readily converted to  coordinate
free setting \cite{car6,tul}. We also consider systems with finite number of
degrees of freedom. This is sufficient to explore the main ideas and
constructions (this can be rendered to a field theory, e.g., using De Witt's notation \cite{dewitt1}).

\section{Legendre transform and multidimensional Clairaut equation}

First, recall the standard Legendre(-Fenchel)  transform for the  theory with nonsingular Lagrangian \cite{arnold}. We then
show its relation to the Clairaut equation \cite{arnold1} in some details
\cite{dup_belg2009}, which will be used to explain the main idea below.
Let\footnote{We use indices in arguments, because by type of index we will
distinguish them below.} $L\left(  q^{A},v^{A}\right)  $, $A=1,\ldots n$, be a
Lagrangian given by a smooth function\footnote{We consider time-independent
case for simplicity and conciseness.} of $2n$ variables ($n$ generalized
coordinates $q^{A}  $ and $n$ velocities $v^{A}  =\dot{q}^{A}  =dq^{A}  /dt$) on the
configuration space $TM$. By definition, a Hamiltonian $H\left(  q^{A},{p}_{A}\right)  $ as a dual function to the Lagrangian (in the second set of
variables ${p}_{A}  $)  constructed by means of
the Legendre(-Fenchel) transform has the form\footnote{We use summation
convention for indices which are not in arguments of functions.}
\begin{equation}
H\left(  q^{A},{p}_{A}\right)  =\sup\limits_{v^{A}}\left[  {p}%
_{B}v^{B}-L\left(  q^{A},v^{A}\right)  \right]  , \label{hqp}%
\end{equation}
where the supremum is taken with fixed $q^{A}  $ and ${p}%
_{A}  $. In doing the Legendre transform, the coordinates
$q^{A}  $ are treated as fixed (passive) parameters of the
duality transformation, and velocities, $v^{A}  $ are \textit{independent} functions of time. Then (\ref{hqp}) leads to the
\textit{supremum condition} %
\begin{equation}
p_{A}=\dfrac{\partial L\left(  q^{A},v^{A}\right)
}{\partial v^{A}}. \label{pl}%
\end{equation}
To obtain a dual
function $H\left(  q^{A},p_{A}\right)  $, we need to get rid of dependence
on velocity in the r.h.s. of (\ref{hqp}). This can be done in \textit{two} ways:

1) \textbf{Direct} way: resolve the condition (\ref{pl}) directly and
obtain its solution as a set of functions $v^{A}=V^{A}\left(  q^{A}%
,p_{A}\right)  $, then substitute them to (\ref{hqp}) and obtain the standard
Hamiltonian on the physical phase space $T^{\ast}M$ (see e.g. \cite{arnold,goldstein1})%
\begin{equation}
H^{st}\left(  q^{A},p_{A}\right)  =p_{B}V^{B}\left(  q^{A},p_{A}\right)
-L\left(  q^{A},V^{A}\left(  q^{A},p_{A}\right)  \right)  . \label{hq}%
\end{equation}
This can be done only in the case of convex Lagrangian function (in the
second
set of variables $v^{A}  $), which is equivalent
to the Hessian being non-zero%
\begin{equation}
\det\left\Vert \dfrac{\partial^{2}L\left(  q^{A},v^{A}\right)  }{\partial
v^{A}\partial v^{B}}\right\Vert \neq0. \label{hs}%
\end{equation}

2) \textbf{Indirect} way: differentiate both sides of (\ref{hs}) by
momenta and use the supremum condition (\ref{pl}) to obtain the
\textquotedblleft dual supremum condition\textquotedblright\ in the form%
\begin{equation}
v^{A}=V^{A}\left(  q^{A},p_{A}\right)=\dfrac{\partial H\left(  q^{A},{p}_{A}\right)  }{\partial{p}%
_{A}}. \label{hv}%
\end{equation}
Then we substitute these velocities to (\ref{hs}), which results in \textit{no
manifest dependence} of $v^{A}$. Thus we obtain a partial differential equation  with respect to Hamiltonian which in fact
is the multidimensional Clairaut equation \cite{dup_belg2009}%
\begin{equation}
H^{cl}\left(  q^{A},\bar{p}_{A}\right)  =\bar{p}_{B}\dfrac{\partial
H^{cl}\left(  q^{A},\bar{p}_{A}\right)  }{\partial\bar{p}_{B}}-L\left(
q^{A},\dfrac{\partial H^{cl}\left(  q^{A},\bar{p}_{A}\right)  }{\partial
\bar{p}_{A}}\right)  . \label{cl}%
\end{equation}

We call the transformation defined by (\ref{cl}) a \textit{Clairaut duality
transform} (or the \textit{Clairaut-Legendre transform}) and
$H^{cl}\left(  q^{A},\bar{p}_{A}\right)  $ a \textit{Clairaut-Hamilton
function}. Note that (\ref{pl}) is normally treated as a \textit{definition} of
dynamical momenta $p_{A}$, but we should distinguish them from the parameters
of the Clairaut duality transform $\bar{p}_{A}$: \textit{before} applying the
supremum condition (\ref{pl}) they are assumed \textit{noncoincidental}. 

The difference between the above two approaches is crucial for
singular Lagrangian theories \cite{car6}. We thus
label the resulting Hamiltonians by different indices.
Specifically, the Clairaut equation (\ref{cl}) \textit{has
solutions} even in the case when the Hessian (\ref{hs}) is zero. So the
Clairaut duality transform is more general and includes the ordinary
duality (Legendre-Fenchel) transform as a particular case. To show this and find
solutions of the Clairaut equation (\ref{cl}), we differentiate it by $\bar
{p}_{A}$ to obtain%
\begin{equation}
\left.  \left[  \bar{p}_{B}-\dfrac{\partial L\left(  q^{A},v^{A}\right)
}{\partial v^{B}}\right]  \right\vert _{v^{A}=\tfrac{\partial H^{cl}\left(
q^{A},\bar{p}_{A}\right)  }{\partial\bar{p}_{A}}}\cdot\dfrac{\partial
^{2}H^{cl}\left(  q^{A},\bar{p}_{A}\right)  }{\partial\bar{p}_{A}\partial
\bar{p}_{B}}=0. \label{plh}%
\end{equation}
So we have two possibilities depending on which multiplier in (\ref{plh}) is zero:

1) Envelope solutions defined by the first multiplier in (\ref{plh})
being zero,  this demand coincides with the supremum condition (\ref{pl}). So we obtain the standard Hamiltonian (\ref{hq})%
\begin{equation}
H_{env}^{cl}\left(  q^{A},\bar{p}_{A}\right)  \vert _{\bar{p}_A=p_A} =H^{st}\left(  q^{A},p_{A}\right)  .
\label{hh}%
\end{equation}

2) A general solution defined the \textquotedblleft dual
Hessian\textquotedblright\ being zero%
\begin{equation}
\dfrac{\partial^{2}H^{cl}\left(  q^{A},\bar{p}_{A}\right)  }{\partial\bar
{p}_{A}\partial\bar{p}_{B}}=0. \label{h0}%
\end{equation}
This gives $\dfrac{\partial H^{cl}\left(  q^{A},\bar{p}_{A}\right)  }%
{\partial\bar{p}_{A}}=c^{A}  $ and then the general
solution acquires the form%
\begin{equation}
H_{gen}^{cl}\left(  q^{A},\bar{p}_{A}\right)  =\bar{p}_{B}c^{B}  -L\left(  q^{A},c^{A}  \right)  ,
\label{gen}%
\end{equation}
where $c^{A}  $ are arbitrary smooth functions considered in the Clairaut equation (\ref{cl}) as parameters. Note
that $H_{gen}^{cl}\left(  q^{A},\bar{p}_{A}\right)  $ is always linear in the
variables $\bar{p}_{A}$ which now do not coincide
with the dynamical momenta, because we do not have the supremum condition
(\ref{pl}).

Now consider a singular Lagrangian theory for which the Hessian (\ref{hs}) is
zero. This means that the rank of Hessian matrix $$W_{AB}=\frac
{\partial^{2}L\left(  q^{A},v^{A}\right)  }{\partial v^{A}\partial v^{B}}$$ is
$r<n$, and we suppose that $r$ is constant. We rearrange indices of
$W_{AB}$ in such a way that a nonsingular minor of rank $r$ appears in the
upper left corner. Represent the index $A$ as follows: if $A=1,\ldots, r$,
we replace $A  $ with $i$ (the ``regular'' index), and, if $A=r+1,\ldots,n$
we replace  $A  $ with $\alpha$ (the ``degenerate'' index). Obviously,
 $\det W_{ij}\neq0$, and $\operatorname{rank}W_{ij}=r$. Thus any set of
variables labelled by a single index splits as a disjoint union of two subsets. We call those subsets
\textit{regular} (having Latin indices) and \textit{degenerate} (having Greek
indices).
\section{Generalized Legendre transform for degenerate Lagrangians}
The standard Legendre transform is not applicable in the singular case
because the condition (\ref{hs}) is not valid \cite{tul}. Therefore the supremum
condition (\ref{pl}) cannot be resolved under degenerate $A$, but it can be
resolved under regular $A$ only, because $\det W_{ij}\neq0$. On the contrary, the
Clairaut duality transform given by (\ref{cl}) independent of the Hessian being zero or not \cite{dup_belg2009}. Thus, we state the main assumption of
the formalism we present here: \textit{the ordinary duality of convex functions can be
generalized to the Clairaut duality for functions with zero Hessian}. This
can be rephrased by saying that the standard
Legendre(-Fenchel) transform of nonsingular Lagrangian theory is generalized to the
Clairaut-Legendre transform, and in both cases the corresponding transformation
 is described by the same Clairaut
equation (\ref{cl}).

To find its solutions, we again differentiate (\ref{cl}) by $\bar{p}_{A}$ and
present the sum (\ref{plh}) in $B$ in two terms: regular and degenerate ones%
\begin{equation}
\left[  \bar{p}_{i}-\dfrac{\partial L\left(  q^{A},v^{A}\right)  }{\partial
v^{i}}\right]  \cdot\dfrac{\partial^{2}H^{cl}\left(  q^{A},\bar{p}_{A}\right)
}{\partial\bar{p}_{A}\partial\bar{p}_{i}}+\left[  \bar{p}_{\alpha}%
-\dfrac{\partial L\left(  q^{A},v^{A}\right)  }{\partial v^{\alpha}}\right]
\cdot\dfrac{\partial^{2}H^{cl}\left(  q^{A},\bar{p}_{A}\right)  }{\partial
\bar{p}_{A}\partial\bar{p}_{\alpha}}=0. \label{pphh}%
\end{equation}

As $\det W_{ij}\neq0$, we suggest to replace (\ref{pphh}) by the conditions%
\begin{align}
&  \bar{p}_{i}=p_{i}=\dfrac{\partial L\left(  q^{A},v^{A}\right)  }{\partial
v^{i}},\label{pp}\\
&  \dfrac{\partial^{2}H^{cl}\left(  q^{A},\bar{p}_{A}\right)  }{\partial
\bar{p}_{A}\partial\bar{p}_{\alpha}}=0. \label{hpp}%
\end{align}

 In this way  we obtain a \textquotedblleft
mixed\textquotedblright\ envelope/general solution of the Clairaut equation,
which can be also treated as a ``partial'' Legendre transform \cite{dup_belg2009}. 

After resolving of (\ref{pp}) under regular velocities $v^{i}=V^{i}\left(
q^{A},p_{i},v^{\alpha}\right)  $ and writing down a solution of (\ref{hpp})  as $$\dfrac{\partial H^{cl}\left(  q^{A},\bar{p}_{A}\right)
}{\partial\bar{p}_{\alpha}}=v^{\alpha} , $$ (where
$v^{\alpha}  $ are arbitrary functions,
unresolved velocities) we obtain a \textquotedblleft mixed\textquotedblright%
\ Clairaut-Hamilton function%
\begin{equation}
H_{mix}^{cl}\left(  q^{A},p_{i},\bar{p}_{\alpha},v^{\alpha}\right)
=p_{i}V^{i}\left(  q^{A},p_{i},v^{\alpha}\right)  +\bar{p}_{\alpha}v^{\alpha
}-L\left(  q^{A},V^{i}\left(  q^{A},p_{i},v^{\alpha}\right)  ,v^{\alpha
}\right)  , \label{hm}%
\end{equation}
which is the desired Clairaut-Legendre transform written in coordinates. Note that (\ref{hm}) coincides with
the \textquotedblleft slow and careful Legendre
transformation\textquotedblright\ of \cite{tul/urb} and with the
\textquotedblleft generalized Legendre transformation\textquotedblright\ of
\cite{cen/hol/hoy/mar}.
\section{Generalized Hamiltonian formalism for singular Lagrangians}
The standard Lagrange equations of motion $\tfrac{d}{dt}\tfrac{\partial
L\left(  q^{A},v^{A}\right)  }{\partial v^{A}}=\tfrac{\partial L\left(
q^{A},v^{A}\right)  }{\partial q^{A}}$ in our notation have the form%
\begin{equation}
\dfrac{dp_{i}}{dt}=\dfrac{\partial L\left(  q^{A},v^{A}\right)  }{\partial
q^{i}},\ \ \ \ \dfrac{dh_{\alpha}\left(  q^{A},p_{i}\right)  }{dt}=\left.
-\dfrac{\partial L\left(  q^{A},v^{A}\right)  }{\partial q^{\alpha}%
}\right\vert _{v^{i}=V^{i}\left(  q^{A},p_{i},v^{\alpha}\right)  }, \label{ph}%
\end{equation}
where%
\begin{equation}
h_{\alpha}\left(  q^{A},p_{i}\right)  =-\left.  \dfrac{\partial L\left(
q^{A},v^{A}\right)  }{\partial v^{\alpha}}\right\vert _{v^{i}=V^{i}\left(
q^{A},p_{i},v^{\alpha}\right)  }. \label{h}%
\end{equation}
The functions $h_{\alpha}\left(  q^{A},p_{i}\right) $ are independent of the unresolved velocities $v^{\alpha}$
since $\operatorname{rank}W_{AB}=r$. One should also take into account that now
$\tfrac{dq^{i}}{dt}=V^{i}\left(  q^{A},p_{i},v^{\alpha}\right)  $ and
$\tfrac{dq^{\alpha}}{dt}=v^{\alpha}$ . Note that before imposing the Lagrange
equations (\ref{ph}) the arguments of
$L\left(  q^{A},v^{A}\right)  $ were treated as independent variables.

A passage to Hamiltonian formalism can be done by the standard procedure:
consider the full differential of both sides of (\ref{hm}) and use the
supremum condition (\ref{pp}), which gives (till now the Lagrange equations
of motion were not used)
\begin{align*}
\dfrac{\partial H_{mix}^{cl}}{\partial p_{i}}  &  =V^{i}\left(  q^{A}%
,p_{i},v^{\alpha}\right)  ,\\
\dfrac{\partial H_{mix}^{cl}}{\partial p_{\alpha}}  &  =v^{\alpha} ,\\
\dfrac{\partial H_{mix}^{cl}}{\partial q^{i}}  &  =-\left.  \dfrac{\partial
L\left(  q^{A},v^{A}\right)  }{\partial q^{i}}\right\vert _{v^{i}=V^{i}\left(
q^{A},p_{i},v^{\alpha}\right)  }+\left[  \bar{p}_{\beta}+h_{\beta}\left(
q^{A},p_{i}\right)  \right]  \dfrac{\partial v^{\beta}}{\partial q^{i}},\\
\dfrac{\partial H_{mix}^{cl}}{\partial q^{\alpha}}  &  =-\left.
\dfrac{\partial L\left(  q^{A},v^{A}\right)  }{\partial q^{\alpha}}\right\vert
_{v^{i}=V^{i}\left(  q^{A},p_{i},v^{\alpha}\right)  }+\left[  \bar{p}_{\beta
}+h_{\beta}\left(  q^{A},p_{i}\right)  \right]  \dfrac{\partial v^{\beta}%
}{\partial q^{\alpha}}.
\end{align*}
An application of the Lagrange equations (\ref{ph}) yields the system of
equations which gives a Clairaut-Hamiltonian description of a singular theory%
\begin{align}
\dfrac{\partial H_{mix}^{cl}}{\partial p_{i}}  &  =\dfrac{dq^{i}}%
{dt},\label{h1}\\
\dfrac{\partial H_{mix}^{cl}}{\partial p_{\alpha}}  &  =\dfrac{dq^{\alpha}%
}{dt},\label{h2}\\
\dfrac{\partial H_{mix}^{cl}}{\partial q^{i}}  &  =-\dfrac{dp_{i}}{dt}+\left[
\bar{p}_{\beta}+h_{\beta}\left(  q^{A},p_{i}\right)  \right]  \dfrac{\partial
v^{\beta}}{\partial q^{i}},\label{h3}\\
\dfrac{\partial H_{mix}^{cl}}{\partial q^{\alpha}}  &  =\dfrac{dh_{\alpha
}\left(  q^{A},p_{i}\right)  }{dt}+\left[  \bar{p}_{\beta}+h_{\beta}\left(
q^{A},p_{i}\right)  \right]  \dfrac{\partial v^{\beta}}{\partial q^{\alpha}}.
\label{h4}%
\end{align}

This system has two disadvantages: 1) It contains the ``nondynamical'' momenta
$\bar{p}_{\alpha}$; 2) It has derivatives
of unresolved velocities $v^{\alpha}$. To get rid of them, we introduce a ``physical''
Hamiltonian%
\begin{equation}
H_{0}\left(  q^{A},p_{i}\right)  =H_{mix}^{cl}\left(  q^{A},p_{i},\bar
{p}_{\alpha},v^{\alpha}\right)  -\left[  \bar{p}_{\beta}+h_{\beta}\left(
q^{A},p_{i}\right)  \right]  v^{\beta}. \label{hph}%
\end{equation}
Using (\ref{pp}) and (\ref{hm}), one can show that the r.h.s. of (\ref{hph})
indeed does not depend on ``nondynamical'' momenta $\bar{p}_{\alpha}$ and
unresolved velocities $v^{\alpha}$. Then from (\ref{h1})--(\ref{h4}) we obtain
the system of (first order differential) equations which describes a singular Lagrangian theory%
\begin{align}
\dfrac{dq^{i}}{dt}  &  =\left\{  q^{i},H_{0}\left(  q^{A},p_{i}\right)
\right\}  +\left\{  q^{i},h_{\beta}\left(  q^{A},p_{i}\right)  \right\}
v^{\beta},\label{q1}\\
\dfrac{dp_{i}}{dt}  &  =\left\{  p_{i},H_{0}\left(  q^{A},p_{i}\right)
\right\}  +\left\{  p_{i},h_{\beta}\left(  q^{A},p_{i}\right)  \right\}
v^{\beta},\label{q2}\\
F_{\alpha\beta}\left(  q^{A},p_{i}\right)  v^{\beta}  &  =D_{\alpha}%
H_{0}\left(  q^{A},p_{i}\right)  , \label{q3}%
\end{align}
where $\left\{  X,Y\right\}  =\tfrac{\partial X}{\partial q^{i}}%
\tfrac{\partial Y}{\partial p_{i}}-\tfrac{\partial Y}{\partial q^{i}}%
\tfrac{\partial X}{\partial p_{i}}$ is the \textquotedblleft
regular\textquotedblright\ Poisson bracket (in regular variables). We introduce here a \textquotedblleft$q^{\alpha}$-long derivative\textquotedblright%
\begin{equation}
D_{\alpha}X=\dfrac{\partial X}{\partial q^{\alpha}}+\left\{  X,h_{\alpha
}\left(  q^{A},p_{i}\right)  \right\}  \label{da}%
\end{equation}
and a \textquotedblleft$q^{\alpha}$-non-Abelian field strength
(curvature)\textquotedblright%
\begin{equation}
F_{\alpha\beta}\left(  q^{A},p_{i}\right)  =\dfrac{\partial h_{\alpha}\left(
q^{A},p_{i}\right)  }{\partial q^{\beta}}-\dfrac{\partial h_{\beta}\left(
q^{A},p_{i}\right)  }{\partial q^{\alpha}}+\left\{  h_{\alpha}\left(
q^{A},p_{i}\right)  ,h_{\beta}\left(  q^{A},p_{i}\right)  \right\}  .
\label{f}%
\end{equation}

The system (\ref{q1})--(\ref{q3}) is equivalent to the Lagrange equations of
motion due to our construction.
\section{New bracket}
In the case $\operatorname{rank}F_{\alpha\beta}\left(  q^{A},p_{i}\right)  =n-r$,
all the velocities $v^{\alpha}$ can be found from (\ref{q3}) in a purely
algebraic way. If $\operatorname{rank}F_{\alpha\beta}\left(  q^{A}%
,p_{i}\right)  =r_{F}<n-r$, then a singular theory has $n-r-r_{F}$ gauge
degrees of freedom. In the first case one can resolve (\ref{q3}) as follows%
\begin{equation}
v^{\beta}=D_{\alpha}H_{0}\left(  q^{A},p_{i}\right)  \bar{F}^{\alpha\beta
}\left(  q^{A},p_{i}\right)  , \label{vd}%
\end{equation}
where $\bar{F}^{\alpha\beta}\left(  q^{A},p_{i}\right)  $ is the inverse matrix to
$F_{\alpha\beta}\left(  q^{A},p_{i}\right)  $ , i.e. $F_{\alpha\beta}\left(
q^{A},p_{i}\right)  \bar{F}^{\beta\gamma}\left(  q^{A},p_{i}\right)
=\delta_{\alpha}^{\gamma}$. Substitute (\ref{vd}) to (\ref{q1})--(\ref{q2})
to present the system of equations for a singular Lagrangian theory
in the Hamiltonian form as follows%
\begin{align}
\dfrac{dq^{i}}{dt}  &  =\left\{  q^{i},H_{0}\left(  q^{A},p_{i}\right)
\right\}  _{F},\label{qh0}\\
\dfrac{dp_{i}}{dt}  &  =\left\{  p_{i},H_{0}\left(  q^{A},p_{i}\right)
\right\}  _{F}, \label{ph0}%
\end{align}
where we define a new bracket%
\begin{equation}
\left\{  X,Y\right\}  _{F}=\left\{  X,Y\right\}  +\left\{  X,h_{\alpha}\left(
q^{A},p_{i}\right)  \right\}  \bar{F}^{\alpha\beta}\left(  q^{A},p_{i}\right)
D_{\beta}Y. \label{xyf}%
\end{equation}

Note that the time evolution of any function $X$ of dynamical variables $\left(
q^{A},p_{i}\right)  $ is also determined by the bracket (\ref{xyf}) as follows%
\begin{equation}
\dfrac{dX}{dt}=\left\{  X,H_{0}\left(  q^{A},p_{i}\right)  \right\}  _{F}.
\label{dx}%
\end{equation}

The bracket (\ref{xyf}) is not anticommutative and does not satisfy Jacobi
identity. Therefore, the standard quantization scheme is not applicable
here. We expect that some more intricate further
assumptions should be made to quantize consistently singular systems within the suggested approach.

\newpage
\section*{Conclusions}
To conclude, we describe Hamiltonian evolution of singular systems
using $n-r+1$ functions of dynamical variables $H_{0}\left(  q^{A}%
,p_{i}\right)  $ and $h_{\alpha}\left(  q^{A},p_{i}\right)  $. This is done by means of the generalized
Clairaut-Legendre transform, that is by solving the corresponding multidimensional
Clairaut equation. All variables are set as regular or
degenerate. We consider the
restricted phase space formed by the regular momenta $p_{i}$ only.

There are two reasons why degenerate momenta $\bar{p}_{\alpha}$ are not
worthwhile to be considered in a singular Lagrangian theory: 

1) the mathematical reason: there is no possibility to
resolve the degenerate velocities $v^{\alpha}$ as can be done for the regular
velocities $v^{i}$ in (\ref{pp}); 

2) the physical reason: momentum is a
\textquotedblleft measure of movement\textquotedblright, but in degenerate
directions there is no dynamics, hence --- no reason to introduce the
corresponding momenta at all. 

Thus there is no notion of constraint \cite{dirac,sundermeyer} as
restriction on ``nondynamical'' momenta, because eventually we do not consider the latter ---
thus nothing to constrain. Under this approach, the degenerate coordinates $q^{\alpha}$
work as parameters analogous to $n-r$ time variables (with $n-r$ corresponding
\textquotedblleft Hamiltonians\textquotedblright\ $h_{\alpha}\left(
q^{A},p_{i}\right)  $, see (\ref{da}) and \cite{lon/lus/pon,dom/lon/gom/pon}). The Hamiltonian form of the equations
of motion (\ref{qh0})--(\ref{ph0}) is achieved by introducing a new bracket
(\ref{xyf}) depending on the above
$n-r+1$ functions. This bracket is responsible for the time evolution. However, is not anticommutative and does not satisfy Jacobi
identity, and therefore its quantization
requires non-Lie algebra methods \cite{myung}.

\medskip

\textit{Acknowledgements}. The author is grateful to G. A. Goldin for kind
hospitality at Rutgers University, where this work was finalized, and to
the
Fulbright Scholar Program for financial support, also he would like to express deep thankfulness to
V.~P.~Aku\-lov, Yu.~A.~Berezhnoj, V. Berezovoj, Yu. Bolotin, B. Broda, V. K. Dubovoy, V. Gershun, U. G\"{u}nter, R. Jackiw, V. D. Khodusov, A.~T.~Kotvytskiy, M. Krivoruchenko, G.~C.~Kurinnoj, M. Lapidus, J. Lukierski, N. Merenkov,
B.~V.~Novik\-ov, A. Nurmagam\-betov, L.~A.~Pastur, S.~A.~Ovsienko, S.~V.~Peletminskij, D. Polyakov, B. Shapiro, W.~Siegel, V. A. Soroka, K.~S.~Stelle, Yu. P. Stepanovsky, P. Urbanski, A. A. Yantzevich, A.~A.~Zheltukhin, M. Znojil and B. Zwiebach for fruitful discussions.
\newpage

\begin{small}

\end{small}

\end{document}